\documentclass[11pt,a4paper]{article}

\usepackage{geometry}
\usepackage{setspace}
\usepackage{authblk}
\usepackage{lmodern}
\usepackage[T1]{fontenc}
\usepackage[english]{babel}
\usepackage{amsmath,amssymb,amsfonts,bm}
\usepackage{graphicx}
\usepackage{url}
\usepackage{hyperref}

\hypersetup{
    colorlinks=true,
    linkcolor=blue,
    citecolor=blue,
    urlcolor=blue,
}

\title{\textbf{Decoherence challenges in Nanoscience: \\ A Quantum Phase Space perspective}}

\author[1]{Ravo Tokiniaina Ranaivoson}
\author[2]{Angelo Mamitiana Ralaikoto}
\author[3]{Diary Lova Ratsimbazafy}
\author[4]{Rivo Herivola Manjakamanana Ravelonjato}
\author[5]{Roland Raboanary}
\author[6]{Raoelina Andriambololona}
\author[7]{Nomenjanahary Tanjonirina Manampisoa}
\author[8]{Fanamby Sahondraniandriana}

\affil[ ]{\footnotesize
  \texttt{tokiniainaravor13@gmail.com}\textsuperscript{1},
  \texttt{ralaikotoangelo@gmail.com}\textsuperscript{2},
  \texttt{ratsimbazafydiary1@gmail.com}\textsuperscript{3},
  \texttt{manjakamanana@yahoo.fr}\textsuperscript{4},
  \texttt{r\_raboanary@yahoo.fr}\textsuperscript{5},
  \texttt{raoelina.andriambololona@gmail.com}\textsuperscript{6},
  \texttt{tanjonaphysics@gmail.com}\textsuperscript{7},
  \texttt{shawinahfana@gmail.com}\textsuperscript{8}
}

\affil[1,2,3,4,6,7,8]{Institut National des Sciences et Techniques Nucl\'eaires, BP 3907 Antananarivo 101, Madagascar}
\affil[1,6,8]{TWAS Madagascar Chapter, Malagasy Academy, BP 4279 Antananarivo 101, Madagascar}
\affil[5]{Faculty of Sciences, University of Antananarivo, BP 566, Antananarivo 101, Madagascar}

\date{}

\begin{document}

\maketitle

\begin{abstract}
Quantum decoherence is both the fundamental mechanism underlying the
quantum-to-classical transition and a major challenge for the
development of scalable nanoscale quantum technologies. This work
introduces a Quantum Phase Space (QPS) framework that provides a unified
geometric description of decoherence. The framework is based on a dual
structure consisting of an overcomplete continuous frame of
minimum-uncertainty states
$\left| \left. \left\langle z \right\rangle \right\rangle \right.\ $
that defines the QPS geometry, and an orthonormal
basis $\{\left| \left. n,\left\langle z(t) \right\rangle \right\rangle \right.\}$
that satisfies the strict mathematical requirements for decoherence
within the Spectrum Broadcast Structure (SBS) objectivity criterion for
pointer states. Within this framework, the variance-covariance matrix
$\mathcal{G}(t)$ of the QPS ground states serves as a universal
indicator of decoherence regimes: it is time-independent
($\dot{\mathcal{G}} = 0$) for Markovian (memoryless) dynamics,
and time-dependent $(\dot{\mathcal{G}} \neq 0)$ for non-Markovian
dynamics with memory and information backflow. To illustrate the
formalism, the Hu-Paz-Zhang (HPZ) model is generalized to
include simultaneous position and momentum couplings to the environment,
leading to a generalized non-Markovian master equation characterized by
a spectral-density matrix. The corresponding evolution equations for the
first
moments$\ (\left\langle p(t) \right\rangle,\left\langle x(t) \right\rangle)$
and for the ground covariance matrix $\mathcal{G(}t)$ establish a direct
connection between microscopic environmental properties, classical-like
trajectories and the QPS geometry. In the Markovian limit, the QPS
geometry becomes time-independent, formalizing the concept that a
stationary environment selects and freezes a specific set of
quasi-classical trajectories, whereas the non-Markovian regime naturally
describes memory effects, information backflow, recoherence,
and evolving deformations of the ground uncertainty ellipse. The proposed framework provides
a unified theoretical foundation for modeling decoherence in nanoscale
systems involving simultaneous position and momentum interactions with
the environment, including nanoelectromechanical resonators,
optomechanical systems, and superconducting quantum circuits. The QPS
framework may thus bridge fundamental theory and practical quantum
engineering, offering a promising coherent pathway to understand,
control, and exploit decoherence at the nanoscience frontier.

\medskip
\textbf{Keywords:} Quantum decoherence, Pointer states, Quantum phase
space, Nanoscience, Quantum technologies.
\end{abstract}

\section{Introduction}

  Quantum decoherence is one of the most fundamental phenomena in modern
physics. It describes the progressive loss of quantum coherence
resulting from the interaction between a quantum system and its
surrounding environment, thereby providing the physical mechanism
underlying the emergence of classical behavior from quantum dynamics.
Beyond its foundational role in the quantum-to-classical transition and
the quantum measurement problem, \cite{Schlosshauer2004,Zurek2003,Schlosshauer2007,Ranaivoson2022b,Wheeler1983,Manampisoa2025,Tomaz2025,Karlsson2025,Breuer2007}, decoherence has become a
central issue in nanoscience and quantum engineering, where it
determines the stability, coherence time, and performance of nanoscale
quantum devices, including quantum computers, quantum sensors,
superconducting circuits, quantum dots, and nanoelectromechanical systems
\cite{Naeij2025,Onizhuk2025,Ladd2010,Lu2025,Tyryshkin2012,Heinrich2021}.

Unlike classical physics, quantum mechanics allows physical systems to
exist in coherent superpositions of different states
\cite{Schlosshauer2004,Zurek2003,Schlosshauer2007,Ranaivoson2022b,Andriambololona1990,Zurek1991}. Nevertheless, the macroscopic world exhibits
well-defined classical properties rather than coherent superpositions.
Quantum decoherence provides the physical mechanism that bridges these
two regimes. Through continuous interaction and entanglement with its
surrounding environment, the reduced density operator of a quantum
system progressively loses its off-diagonal coherence terms and becomes
approximately diagonal in a preferred basis of so-called pointer
states. These environmentally selected states are comparatively robust
against decoherence, remaining stable under the system-environment
interaction and thereby giving rise to the emergence of classical-like
behavior \cite{Schlosshauer2004,Zurek2003,Schlosshauer2007,Ranaivoson2022b, Wheeler1983,Manampisoa2025,Zurek1991,Brasil2015,Duruisseau2023,Chisholm2026,Zurek2022,Feller2020,Zurek1993a,Zurek1993b,Dalvit2005}.

During the past decades, the characterization of pointer states has been
driven by complementary paradigms such as environment-induced
superselection (einselection) \cite{Zurek2003,Zurek2022,Feller2020}, the Predictability
Sieve criterion \cite{Zurek1993a,Zurek1993b,Dalvit2005}, Quantum Darwinism \cite{Duruisseau2023,Chisholm2026,Zurek2022,Zurek2009},
and the Spectrum Broadcast Structure (SBS) \cite{Horodecki2015,Korbicz2021}. Despite the
significant advancements these theories offer, developing a unified and
mathematically rigorous framework for decoherence, one that, in
particular, accurately captures the identification and dynamical
evolution of pointer states within realistic environments, remains an
open challenge \cite{Brasil2015,Duruisseau2023}. In parallel, understanding the role of
environmental memory has become increasingly important across quantum
physics, from fundamental studies to nanoscale technologies and quantum
devices \cite{Breuer2007,Onizhuk2025,Ladd2010,Lu2025,Tyryshkin2012,Heinrich2021,Breuer2009,deVega2017,Breuer2016,Malekakhlagh2016,Lacroix2024,Brandner2025,Odeh2025}. Although Markovian dynamics
is now widely recognized as a particular limit of the more general
non-Markovian description, a unified geometric framework capable of
relating both regimes to the evolution of quantum states is still
lacking. In realistic nanoscale systems, finite environmental memory,
information backflow, and strong system-environment correlations
profoundly influence decoherence \cite{Malekakhlagh2016,Lacroix2024,Brandner2025,Odeh2025,Ranaivoson2022}, making it desirable to
develop a formalism in which both Markovian and non-Markovian dynamics
emerge naturally from the same underlying geometric description.

In this work, we address these challenges by introducing a novel
theoretical framework rooted in the recently developed concept of
Quantum Phase Space (QPS) \cite{Ranaivoson2022,Ravelonjato2023,Ranaivoson2025}. The proposed approach provides
a unified and mathematically rigorous description of quantum decoherence
based on a geometric representation of quantum states in phase space.
The QPS formalism is characterized by a dual mathematical structure: an
overcomplete continuous frame of joint momentum-coordinate
minimum-uncertainty states, denoted
$\left\{ \left| \left. \ \left\langle z \right\rangle \right\rangle \right.\  \right\}$,
analogous to the family of states known as coherent and squeezed states
in the literature \cite{Ranaivoson2022b,Manampisoa2025,Ranaivoson2022,Philbin2014,Bagchi2020,Dodonov2002,RosasOrtiz2019,Dey2018}, and defines the geometric
structure of the QPS; and an associated orthonormal basis
$\left\{ \left| n,\left. \ \left\langle z(t) \right\rangle \right\rangle \right.\  \right\}$
which satisfies the strict mathematical requirements expected of a
pointer-state basis. This construction establishes a natural link
between quantum phase space geometry, environment-induced microscopic
dynamics, and the emergence of classical-like behavior, ultimately
paving the way for a unified geometric framework to describe quantum
decoherence.

Leveraging this dual mathematical structure, we show that the QPS serves
as a natural point of convergence for all major pointer-state criteria.
Under this framework, Einselection, Quantum Darwinism , and Spectrum
Broadcast Structure (SBS) collectively identify the orthonormal basis
$\left\{ \left| n,\left.\left\langle z(t) \right\rangle \right\rangle \right.\right\}$
as the preferred pointer basis. Concurrently, criteria focused on
dynamical stability, such as the Predictability Sieve and minimal
entropy production directly single out the "ground state"
$\left| \left. \left\langle z \right\rangle \right\rangle \right.\ $
as the most classically predictable pure state.

To illustrate the power of this geometric framework and bridge it with
established methodologies for microscopic open-system dynamics, we
revisit the Hu-Paz-Zhang (HPZ) master equation, one of the most
comprehensive non-Markovian descriptions of quantum Brownian motion
\cite{Hu1992,Halliwell1996,Homa2023,Homa2020}. Generalizing the interaction Hamiltonian to encompass
simultaneous position and momentum couplings between the system and its
environment naturally introduces a spectral-density matrix. This matrix
supersedes the conventional scalar spectral density, ultimately yielding
a generalized non-Markovian master equation. From this generalized
master equation, we obtain explicit closed-form evolution equations for
the expectation values pair
$(\left\langle p(t) \right\rangle,\left\langle x(t) \right\rangle)$ and
the covariance matrix $\mathcal{G}(t)$, directly linking the microscopic
properties of the environment to the geometry of the QPS. This
generalized framework provides a suitable foundation for modeling
decoherence in nanoscale systems where both position and momentum
interactions with the environment play a significant role.

The paper is structured as follows. Section 2 reviews the foundational
principles of quantum decoherence and the role of pointer states in
multi-level quantum systems, presenting the four main selection
criteria. Section 3 examines the implications of decoherence for
particle motion in nanoscience and quantum technologies, emphasizing the
distinction between Markovian and non-Markovian regimes. Section 4
introduces the quantum phase space framework, establishes the
overcomplete continuous frame and the associated time-dependent
orthonormal basis, and identifies potential pointer states for particle
motion. Section 5 provides an illustrative application of the formalism
by presenting the standard HPZ model and extending it to include
momentum coupling, deriving the generalized master equation and the
explicit evolution equations for expectation values and the ground
variance-covariance matrix. Finally, we conclude with a discussion of
the broader implications of our approach for quantum technologies,
fundamental physics, and the relativistic and cosmological extensions of
the QPS framework. Throughout this paper, quantum operators are denoted
in boldface to distinguish them from their classical counterparts and
expectation values.

\section{Quantum decoherence and pointer states}

\subsection{General Formalism}

Pointer states are the quantum states that remain stable when the system
interacts with its environment. They are the states that decoherence
does not "destroy" \cite{Schlosshauer2004,Zurek2003,Schlosshauer2007,Ranaivoson2022b,Wheeler1983,Manampisoa2025,Tomaz2025,Karlsson2025,Breuer2007}. Decoherence tends to destroy
superpositions through environmental entanglement, but pointer states
are the ones that do not become entangled in a way that destroys their
classical‑like behavior. They are the states that survive environmental
monitoring. We can think of them as the eigenstates of the system
observables that are continuously "measured" by the environment
\cite{Schlosshauer2004,Zurek2003,Schlosshauer2007,Ranaivoson2022b,Manampisoa2025,Zurek2022,Feller2020}. A first way to technically define pointer
states is then to consider them as the eigenstates (or approximate
eigenstates) of the system operators that the environment continuously
monitors. Because the environment acts like a measuring device, it
selects a preferred basis in which superpositions rapidly decay. This
basis is formed by the set of pointer states and is then called the
pointer basis \cite{Schlosshauer2004,Zurek2003,Schlosshauer2007,Zurek2022}. Suppose the interaction Hamiltonian
is:

\begin{equation}
\mathbf{H}_{I} = \ \sum_{k}^{}{\mathbf{A}_{{k}} \otimes \mathbf{B}_{{k}}}
\end{equation}

where operators $\mathbf{A}_{{k}}$ act on the system and
$\mathbf{B}_{{k}}$ on the environment. Pointer states
$\left| \left. \ s_{n} \right\rangle \right.\ $ satisfy:

\begin{equation}
\mathbf{A}_{{k}}\left| s_{n} \right\rangle \approx a_{nk}\left| s_{n} \right\rangle
\end{equation}

The relation (2) means that pointer states are considered as approximate
eigenstates of the operators $\mathbf{A}_{{k}}$ that couple to
the environment:

\begin{itemize}
\item For an environment that measures momentum, the pointer states are
  approximately momentum eigenstates.

\item If the environment measures position, then pointer states are
  localized position states.

\item If it measures energy, pointer states are energy eigenstates.

\item If it measures a spin axis, pointer states are spin states along that
  axis, etc.
\end{itemize}

The evolution of the density operator offers another viewpoint for
defining and visualizing both the emergence of pointer states and the
decoherence phenomenon. Consider the system starting in a pure state
$\left| \left.\psi_{S} \right\rangle \right.$ and the environment
in an initial state $\left| \left. e_{0} \right\rangle \right.\ $.
Suppose the interaction between the system and environment begins at
this initial time $(t = 0)$. For $t > 0$, entanglement arises between
the state of the system and that of the environment. The entangled
global state $\left| \left.\Psi_{SE} \right\rangle \right.\ $ can be
written in the following form:

\begin{equation}
\left| \psi_{SE} \right\rangle = \sum_{n}^{}{C^{n}\left( \left| \left. s_{n} \right\rangle \right.\otimes \left| e_{n} \right\rangle \right)}
\end{equation}

where {$\left| \left.s_{n} \right\rangle \right.\ $} is a basis in
the system state space and $\left| e_{n} \right\rangle$ the basis of the
environment state space correlated with
{$\left| \left. s_{n} \right\rangle \right.\ $}. The canonical density
operator which describes the system‑environment entanglement is then:

\begin{equation}
\mathbf{\rho}_{SE} = |\Psi_{SE}\rangle\langle\Psi_{SE}| = \sum_{n}^{}{\sum_{m}^{}{C^{n}C^{m*}}(|s_{n}\rangle\langle s_{m}| \otimes |e_{n}\rangle\langle e_{m}|)}
\end{equation}

The density operator $\mathbf{\rho}_{S} = \mathbf{\rho}$ of the system
can be obtained by taking the partial trace over the environment degree
of freedom: $\mathbf{\rho} = \operatorname{Tr}_{E}\left( \mathbf{\rho}_{SE} \right)$.
Explicitly:

\begin{equation}
\mathbf{\rho} = \sum_{k}^{}\left\langle e_{k}\left| \mathbf{\rho}_{{SE}} \right|e_{k} \right\rangle = \sum_{n}^{}{\sum_{m}^{}\rho_{m}^{n}\left| s_{n} \right\rangle\left\langle s_{m} \right|}
\end{equation}

With

\begin{equation}
\rho_{m}^{n} = \ \sum_{k}^{}{C^{n}C^{m*}\left( \left\langle e_{k} \middle| e_{n} \right\rangle\left\langle e_{m} \middle| e_{k} \right\rangle \right)} = \left\langle s_{n} \middle| \mathbf{\rho} \middle| s_{m} \right\rangle
\end{equation}

$\{|s_{n}\rangle\}$ is the pointer basis of the system if, for the
environment basis $\{|e_{n}\rangle\}$ correlated with it, we have the
following relation:

\begin{equation}
\left\langle e_{n} \middle| e_{m} \right\rangle \simeq \delta_{m}^{n} = \left\{ \begin{array}{ll}
1 & \text{if } n = m \\
0 & \text{if } n \neq m
\end{array} \right. \quad \text{for} \quad t \gg \tau_{D}
\end{equation}

in which $\tau_{D}$ is called the decoherence time: it is the
characteristic time scale after which decoherence effects become
significant. We have then

\begin{equation}
\rho_{m}^{n}(t) \simeq \sum_{n}^{}{\sum_{m}^{}{{C^{n}C}^{m*}\delta_{n}^{k}\delta_{k}^{m}}} = \left\{ \begin{array}{ll}
\left| C^{n} \right|^{2} & \text{if } n = m \\
0 & \text{if } n \neq m
\end{array} \right. \quad \text{for} \quad t \gg \tau_{D}
\end{equation}

which is equivalent to writing for the system's density operator:

\begin{equation}
\mathbf{\rho}(t) \simeq \sum_{n}^{}{\rho_{n}^{n}\ \left| s_{n} \right\rangle\left\langle s_{n} \right|} = \sum_{n}^{}{\left| C^{n} \right|^{2}\left| s_{n} \right\rangle\left\langle s_{n} \right|} \quad \text{for} \quad t \gg \tau_{D}
\end{equation}

Relations (7), (8), and (9) are equivalent and describe the phenomenon
of decoherence, which consists of a rapid decay of the off‑diagonal
terms of the density matrix for $t > \tau_{D}$. Indeed, these terms
represent the existence of coherent superpositions among pointer states.
For $t \gg \tau_{D}$, the matrix representation of the density operator
in the pointer basis becomes approximately diagonal. The density
operator then becomes an incoherent combination (statistical mixture) of
pointer states: the diagonal elements $\rho_{n}^{n} = \left| C^{n} \right|^{2}$ simply represent
classical‑like probabilities of finding the system in a defined pointer
state.

\subsection{Pointer-State selection criteria}

The identification of the pointer basis is a central task in the theory of decoherence. While the general formalism presented in the previous section provides the conceptual framework, several independent criteria have been developed to determine the pointer states rigorously. However, these criteria should be considered complementary, as they approach the problem from different physical perspectives: structural, dynamical, and informational. Moreover, as we will show in Section 4, they converge within the Quantum Phase Space (QPS) framework. The main criteria are outlined below.

\subsubsection{Environment‑induced superselection (Einselection)}

Einselection \cite{Zurek2003,Zurek2022,Feller2020}, introduced by Zurek, posits that the environment continuously monitors a set of observables of the system through the interaction Hamiltonian. The pointer states are the approximate eigenstates of the system operators that couple to the environment. This criterion corresponds to the one previously considered in the preceding section to define pointer states via relations (1) and (2). Assuming the interaction Hamiltonian is given by relation (1), the pointer states $\left| \mathbf{s}_{\mathbf{n}} \right\rangle$ are those that satisfy the condition that they are eigenstates of the operators $\mathbf{A}_{{k}}$ in the limit where the self‑Hamiltonian of the system is negligible. More generally, when neither the self‑Hamiltonian nor the interaction dominates, the pointer states arise from a compromise between self‑evolution and interaction, and they become localized in phase space. In the opposite limit, where the dynamics is dominated by the system's self‑Hamiltonian, einselection produces pointer states that coincide with the energy eigenstates of the self‑Hamiltonian. Thus, the environment acts as a measuring apparatus that continuously monitors the observables. States that are not eigenstates of these observables become entangled with the environment and lose coherence. The pointer states, by contrast, remain stable because they are minimally perturbed by the monitoring process. This criterion selects the pointer basis structurally, based on the form of the interaction and the relative strength of the self‑Hamiltonian.

\subsubsection{Predictability Sieve (Minimal entropy production)}

The Predictability Sieve \cite{Zurek1993a,Zurek1993b,Dalvit2005}, introduced by Zurek, Habib, and Paz takes a dynamical perspective.  It identifies pointer states as the pure states that minimize the loss of predictability, that is, the increase in von Neumann entropy, during the decoherence process. For an initial pure state $\left| \left. \mathbf{\psi} \right\rangle \right.$, let $\mathbf{\rho}\left({t} \right)$ be its reduced density matrix after evolution to time ${t}$ ($\mathbf{\rho}\left( {0} \right) = \left| \left.\mathbf{\psi} \right\rangle\right.\left\langle \left.\mathbf{\psi} \right| \right.\ $). The von Neumann entropy

\begin{equation}
\mathcal{S}(t) = - \operatorname{Tr}\lbrack\mathbf{\rho}(t)\ln\mathbf{(\rho}\left({t} \right){)\rbrack}
\end{equation}

measures the loss of purity due to entanglement. The pointer states are selected by the minimization of this entropy over all initial pure states with the requirement that the answer be robust when varying $t$ within a reasonable range. Equivalently, one can minimize the \text{decrease of purity} ${P = Tr\lbrack}{\mathbf{(}\mathbf{\rho}\mathbf{)}}^{\mathbf{2}}\mathbf{\rbrack}$ since the loss of purity is directly related to the increase of von Neumann entropy. This criterion asks: "If I prepare the system in a pure state, which initial state remains the most predictable (i.e., least entangled with the environment) after decoherence?" The answer selects the states that produce the least entropy or, equivalently, lose the least purity. It is a  dynamical criterion that depends on the full evolution, including the system Hamiltonian and the bath correlation functions.

\subsubsection{Quantum Darwinism}

Quantum Darwinism \cite{Duruisseau2023,Chisholm2026,Zurek2022,Zurek2009}, developed by Zurek, explains the emergence of classical objectivity through a Darwinian mechanism: the environment selectively proliferates information about certain system observables, creating multiple redundant copies. The pointer states are those whose information is  redundantly encoded in the environment, allowing multiple observers to independently infer the system's state without perturbing it. The term "Darwinism" reflects the analogy to biological natural selection: pointer states are the "fittest" states, they survive environmental monitoring and leave numerous "descendants" by imprinting redundant copies of their information onto environmental fragments. The hallmark of Quantum Darwinism is the mutual information plateau: for any sufficiently large environment fragment ${F}$, the quantum mutual information between the system ${S}$ and the fragment ${F}$ is nearly equal to the von Neumann entropy of the system:

\begin{equation}
\mathcal{I}(S:F) \simeq \mathcal{S}(S)
\end{equation}

Here, $\mathcal{I}(S:F) = \mathcal{S}(S) + \mathcal{S}(F) - \mathcal{S}(SF)$ is the quantum mutual information, which measures the total correlations between the system and the fragment. The entropy of the system is $\mathcal{S}(S) = - \operatorname{Tr}\mathbf{\lbrack}\mathbf{\rho}_{{S}}\ln\mathbf{(}\mathbf{\rho}_{{S}}{)\rbrack}$. The plateau indicates that the fragment contains almost all the information about the system that is available in the whole environment. The pointer basis is the basis in which the system's state becomes approximately diagonal after decoherence:

\begin{equation}
\mathbf{\rho}_{{S}} \ \mathbf{\simeq}\sum_{{n}}^{}{{p}_{{n}}\left| {s}_{{n}} \right\rangle\left\langle {s}_{{n}} \right|}
\end{equation}

The environment records multiple, independent copies of information about the system's pointer states. This redundancy ensures that different observers who measure different parts of the environment will agree on the system's state, a hallmark of classical objectivity. This criterion is informational: it quantifies how much information about the system is present in environmental fragments. The correlations between the system and the environment may retain some quantumness (quantum discord) that prevents perfect local readout \cite{deVega2017}. Thus, Quantum Darwinism provides a sufficient condition for the emergence of objectivity but not a necessary structural one.

\subsubsection{Spectrum Broadcast Structure (SBS)}

Spectrum Broadcast Structure \cite{Horodecki2015,Korbicz2021}, formalized by R. Horodecki, J. K. Korbicz, and P. Horodecki and further developed by Korbicz, provides a stronger, structural formulation of objectivity. While Quantum Darwinism focuses on the amount of redundant information (mutual information plateau), SBS specifies the form that this redundancy must take to ensure perfect classical readout. A state is said to exhibit SBS when the environment can be partitioned into fractions, each containing sufficient locally accessible information to perfectly distinguish the system pointer states. The global state $\mathbf{\rho}_{{S}{E}_{{1}}{\ldots}{E}_{{N}}}$ of the system and $N$ environment fragments ${E}_{{1}},\ldots,{E}_{{N}}$, has SBS if it takes the specific tensor product form

\begin{equation}
\mathbf{\rho}_{{S}{E}_{{1}}{\ldots}{E}_{{N}}}{=}\sum_{{n}}^{}{p}_{{n}}\left| \left.\mathbf{s}_{{n}} \right\rangle \right.\left\langle \left.\mathbf{s}_{{n}} \right| \right.\mathbf{\otimes}\mathbf{\rho}_{{E}_{{1}}}^{\left({n} \right)}\mathbf{\otimes \ldots \otimes}\mathbf{\rho}_{{E}_{{N}}}^{\left({n} \right)}
\end{equation}

Equivalently, for any environment fragment ${E}_{{k}}$, the reduced state is

\begin{equation}
\mathbf{\rho}_{\mathbf{S}{E}_{{k}}}{=}\sum_{{n}}^{}{p}_{{n}}\left| \left. {s}_{{n}} \right\rangle \right. \left\langle \left.{s}_{{n}} \right| \right.{\otimes}\mathbf{\rho}_{{E}_{{k}}}^{\left({n} \right)}
\end{equation}

Where $\mathbf{\{}\left| \left. \ \mathbf{s}_{{n}} \right\rangle \right.\ \mathbf{\}}$ is an orthonormal pointer basis satisfying $\left\langle s_{n} \middle| s_{m} \right\rangle = \delta_{nm}$, and the conditional states ${\rho}_{{E}_{{k}}}^{\left( {n} \right)}{\ (k = 1,\ldots,N)}$ of each environment fragment satisfy

\begin{equation}
{Tr}\left\lbrack \mathbf{\rho}_{{E}_{{k}}}^{\left({n} \right)}{\rho}_{{E}_{{k}}}^{\left({m} \right)} \right\rbrack{= 0\ for\ n \neq m}
\end{equation}

This condition ensures that an observer accessing a single environment fragment can unambiguously identify the system's pointer state. Consequently, each environmental fragment independently carries complete and locally accessible classical information about the pointer state. The global state exhibits a broadcast structure in which the classical information associated with the system's pointer state is redundantly encoded in multiple, conditionally independent environment fragments. As a result, multiple observers can independently infer the system's pointer state by measuring their respective environment fragments without disturbing either the system or one another, and they will necessarily reach the same conclusion. Spectrum Broadcast Structure therefore provides a rigorous criterion for objective classical reality. Moreover, every SBS state satisfies the redundancy criterion underlying Quantum Darwinism, whereas the converse is not generally true: redundancy of mutual information alone does not guarantee the existence of an SBS.

\section{Particle motion decoherence in nanoscience and quantum technologies}

Quantum decoherence in particle motion refers to the loss of quantum coherence in the spatial or motional degrees of freedom of a particle due to environmental interactions \cite{Schlosshauer2004,Zurek2003,Schlosshauer2007,Ranaivoson2022b,Manampisoa2025,Heinrich2021,Ekinci2005,Seoanez2007}. At the nanoscale (approximately between $1\, nm = 10^{-9}\, m$ and $100\, nm = 10^{-7}\, m$), this process fundamentally determines whether particle motion exhibits quantum behavior (superposition, interference) or classical trajectories. The distinction between Markovian (memoryless environment) and non‑Markovian (memory‑retaining environment) decoherence processes is critical for accurately modeling, controlling, and harnessing quantum motion in nanodevices \cite{Breuer2009,deVega2017,Breuer2016,Malekakhlagh2016,Lacroix2024,Brandner2025,Odeh2025,Schafer2024,Khan2025}.

\subsection{Markovian and Non‑Markovian Regimes in nanoscale particle motion decoherence}

In nanoscience, particle motions, whether of electrons, excitons,
nanoparticles, or nanomechanical resonators, are governed by quantum
mechanics but are highly susceptible to decoherence \cite{Heinrich2021,Odeh2025,Ekinci2005,Seoanez2007,Schafer2024,Khan2025,Ishizaki2009}. The spatial superposition of a particle's wavefunction
decays due to coupling with environmental degrees of freedom (phonons,
photons, charge fluctuators, defect, impurities, etc.) \cite{Breuer2007,Heinrich2021,Odeh2025,Seoanez2007,Schafer2024,Suter2016}. The timescale and nature of this decay
depend critically on whether the environment is Markovian or
non‑Markovian \cite{Breuer2009,deVega2017,Breuer2016,Odeh2025,Schafer2024,Suter2016}. Understanding these
regimes guides material choice and device design.

\begin{itemize}
\item \textbf{Markovian decoherence in particle motion:} under conditions of weak
  coupling, high temperature, and a broadband environment (e.g.,
  electron‑phonon scattering in mesoscopic conductors), decoherence is
  approximately memoryless i.e. Markovian. The off‑diagonal elements of
  the reduced density matrix for a particle in a spatial superposition
  of separation $(\Delta x = |x - x'|)$ decay exponentially:

\begin{equation}
\rho(x,x',t) = \left\langle x\left| \rho(t) \right|x' \right\rangle = \rho(x,x',0)e^{- \left\lbrack \Gamma \cdot t \cdot (\Delta x)^{2} \right\rbrack}
\end{equation}

where $\Gamma = \frac{2M\gamma k_{B}T}{\hslash^{2}}$ is the constant
decoherence rate, $M$ is the particle mass, $\gamma$ is the friction
coefficient, $k_{B}$ is the Boltzmann constant and $T$ is the
temperature. This Markovian description applies, for instance to:

\begin{itemize}
\item Electron transport in quantum dots at $T > 1K$, where
  electron‑electron scattering provides rapid environmental relaxation.
\item Nanomechanical resonators in low‑pressure gas environments
  $(P \sim 1 - 100\, Pa)$ where frequent gas collisions create
  memoryless damping \cite{Zurek1991}.
\end{itemize}

\item \textbf{Non‑Markovian decoherence in particle motion:} In contrast, when
  the environment possesses a structured spectrum, finite correlation
  times, or strong system-environment coupling, decoherence becomes
  non-Markovian. The off-diagonal elements of the density matrix in the
  position representation are then described by

\begin{equation}
\rho(x,x',t) = \left\langle x\left| \rho(t) \right|x' \right\rangle = \rho(x,x',0)e^{- \left\lbrack \Gamma(t) \cdot (\Delta x)^{2} \right\rbrack}
\end{equation}

In contrast to Eq. (16), the decoherence function $\Gamma(t)$ in
equation (17) is time dependent and need not increase monotonically. Its
non-monotonic behavior may lead to partial recoherence, reflecting the
backflow of quantum information from the environment to the system,
which is a characteristic of non-Markovian dynamics.
\end{itemize}

\subsection{Quantum technologies and motional qubits}

Quantum technologies exploit coherent particle motion for information
processing, sensing, and communication (motional qubits). The coherence
time of motional qubits, whether charge, phononic, or mechanical, is
ultimately limited by decoherence, whose Markovian or non‑Markovian
character dictates control strategies. Markovian regimes permit simple
exponential models and conventional dynamical decoupling. Non‑Markovian
regimes require advanced techniques: reservoir engineering, tailored
pulse sequences that synchronize with backflow, and error correction
adapted to non‑exponential decay.

\begin{itemize}
\item \textbf{Charge qubits in quantum dots:} A charge qubit is encoded in the
  orbital occupancy of a single electron shared between two coupled
  quantum dots. Decoherence is primarily dominated by charge noise,
  while electron--phonon interactions contribute mainly to energy
  relaxation. At sufficiently high temperatures, or when the
  environmental correlation time is much shorter than the qubit
  dynamics, decoherence is commonly modeled using the Markov
  approximation, leading to approximately exponential dephasing
  $(\tau_{D} \simeq 10^{-9}s)$ \cite{Heinrich2021,Hayashi2003}. However,
  at low temperatures $(T < 100\, mK)$, the noise spectrum becomes
  structured, revealing non‑Markovian signatures that must be accounted
  for in high‑fidelity control \cite{Odeh2025,Suter2016,Chiang2021,Lastra2011,Ablimit2021}.

\item \textbf{Nanomechanical qubits:} a nanomechanical resonator's center‑of‑mass
  motion can, in principle, be prepared in a spatial superposition.
  Decoherence arises from coupling to thermal phonons, radiation, and
  defects. In the Markovian regime (e.g., moderate vacuum,
  $T > 100\, mK$), energy decay is exponential and described by a
  constant quality factor $Q$. In the non‑Markovian regime (cryogenic,
  high‑strain conditions), interaction with two‑level systems leads to
  complex decay profiles and memory effects, impacting sensing
  performance and quantum state lifetime \cite{Ekinci2005,Seoanez2007,Schafer2024,Joo2024,Engelsen2024,Savel2007,Cleland2024,Bozkurt2025}.
\end{itemize}

\subsection{Challenges related to decoherence in particle motion}

Understanding and controlling decoherence is pivotal for exploiting the
quantum properties of nanoscale particle motion. The journey from
fundamental characterization to practical mitigation involves navigating
a series of interconnected challenges, particularly when environments
exhibit memory (non‑Markovian dynamics). These challenges span
experiment, engineering and theory defining the current frontier of the
field:

\begin{itemize}
\item \textbf{Characterizing decoherence regimes:} determining whether particle
  motion in a given nanoscale device undergoes Markovian or
  non‑Markovian decoherence is nontrivial. Characterization typically
  relies on measurements of coherence dynamics, environmental
  correlation times, and spectral properties, together with theoretical
  reconstruction of system-environment correlations
  \cite{Breuer2009,deVega2017,Breuer2016,Rivas2010,Breuer2012} as well as employing coherence witnesses
  to detect the survival of quantum coherence in these complex
  environments \cite{Li2012}.

\item \textbf{Modeling and simulation}: non-Markovian dynamics are
  computationally demanding to simulate. Developing efficient numerical
  methods, such as the reduced hierarchy equation approach and
  non-Markovian quantum jump techniques, is essential for accurate
  device prediction \cite{Lacroix2024,Ishizaki2009,Piilo2008}.

\item \textbf{Mitigation in non‑Markovian environments}: while Markovian
  decoherence may often be reduced using dynamical decoupling, reservoir
  engineering, feedback cooling, or, where applicable, quantum
  error-correction techniques, non‑Markovian decoherence requires
  environment‑aware strategies \cite{Heinrich2021,Suter2016}: engineering bandgaps to
  suppress emission, using defect‑free materials, or exploiting memory
  effects for enhanced sensing and quantum control.

\item \textbf{Harnessing non‑Markovianity}: A paradigm shift is emerging: rather
  than suppressing all decoherence, non-Markovian memory effects and the
  associated backflow of quantum information can be harnessed to achieve
  quantum advantage \cite{Lacroix2024,Brandner2025,Chiang2021,Fanchini2014}. Structured environments
  can be engineered to enhance quantum performance or enable
  spatiotemporal control of quantum devices.

\item \textbf{Scalability with motional qubits:} as quantum devices scale,
  cross‑talk between motional degrees of freedom and collective
  decoherence channels may become increasingly important. Managing these
  in both Markovian and non‑Markovian contexts is critical for
  multi‑qubit processors and sensor arrays \cite{Onizhuk2025,Heinrich2021,Lacroix2024,Wu2025}.

\item \textbf{Mathematical‑physical framework for pointer states description:} A
  major theoretical and practical challenge is the identification of
  pointer states. In particle motion, these correspond to localized
  positions or well-defined motional states that are least susceptible
  to environmental decoherence \cite{Zurek1991,Brasil2015,Zurek1993a,Zurek1993b,Dalvit2005}. A rigorous
  mathematical framework capable of describing environment-induced
  superselection and the emergence of pointer states in both Markovian
  and non-Markovian regimes would provide a unified foundation for
  understanding decoherence. Such a framework would directly address the
  challenges of characterization, mitigation, and scalability by
  establishing a systematic roadmap for preserving quantum coherence in
  nanoscale particle motion. In the following sections, we show that an
  approach based on Quantum Phase Space (QPS) theory offers a promising
  route toward this objective by providing a single geometric
  representation in which coherence decay and pointer-state selection
  arise as complementary aspects of the same underlying dynamics.
\end{itemize}

\section{Quantum Phase Space perspective on decoherence}

The challenges outlined in the previous sections, notably the identification of pointer states and the characterization of Markovian and non-Markovian decoherence, highlight the need for a unified theoretical framework. In this section, we introduce a formalism rooted in the concept of Quantum Phase Space (QPS), which addresses these challenges by providing a geometric and mathematically rigorous description of decoherence in particle motion.

\subsection{Pointer States for particle motion}

According to decoherence theory, when a particle's motion is subjected
to environmental effects, the interaction results in a selection of
pointer states $\left| \left. \alpha \right\rangle \right.\ $ that
form the pointer basis. The system's instantaneous density operator
$\mathbf{\rho}(t)$ can be decomposed in this basis.

An explicit description of decoherence hinges on a key unknown: the set
of pointer states that are selected by the environment. Physically,
these should correspond to the most stable, quasi‑classical descriptions
of the particle's trajectory. Consequently, the search naturally focuses
on the momentum eigenstates $\left| \left.p \right\rangle \right.\ $,
the position eigenstates $\left| \left. x \right\rangle \right.\ $,
(as suggested by the decoherence mechanisms described in Eqs. (16) and
(17)), or hybrid states, such as coherent states, that combine both
position and momentum localization \cite{Philbin2014,Bagchi2020,Dodonov2002,RosasOrtiz2019,Dey2018}. Indeed, the
interaction Hamiltonian describing the coupling between a system and its
environment commonly involves the position and/or momentum operators,
making these observables natural candidates for environment-induced
superselection \cite{Schlosshauer2007,Zurek1991,Zurek2022}. In this sense, the environment acts
as a continuous monitoring device through position-sensitive
interactions and momentum-dependent scattering processes. Crucially, any
rigorous selection or identification of pointer states must incorporate
the constraints of the uncertainty principle. A rigorous framework for
selecting pointer states must explicitly respect the quantum uncertainty
relation, which is

\begin{equation}
\left( \sigma_{pp} \right)\left( \sigma_{xx} \right) - \left( \sigma_{px} \right)^{2} \geq \frac{(\hslash)^{2}}{4}
\end{equation}

where $\sigma_{pp}$, $\sigma_{xx}$ and $\sigma_{px}$ are respectively
variances and covariance of momentum and coordinate. Selecting a pure
momentum eigenstate $\left| \left. p \right\rangle \right.\ $ implies
$\sigma_{pp} = 0$, which is unphysical as it requires
$\sigma_{xx} = \infty$. Selecting a pure position eigenstate
$\left| \left. x \right\rangle \right.\ $ implies $\sigma_{xx} = 0$,
which is equally unphysical as it requires $\sigma_{pp} = \infty$. If we
rigorously take this conclusion into account, it is clear that even the
standard exponential decay expressions (16) and (17) are ill‑defined
(correspond to idealizations that neglect the fundamental minimum
uncertainty of quantum states).

Therefore, viable pointer states must be states that optimally localize
both position and momentum within the bounds of the uncertainty
principle. Furthermore, in classical physics, the one‑dimensional motion
of a particle is fully described by its instantaneous position $x(t)$
and momentum $p(t)$. The set of all possible simultaneous pairs
$(p,x)$ defines the particle's classical phase space, and each such pair
represents an elementary basis state of motion. It follows that the
quantum pointer states $\left| \left. \alpha \right\rangle \right.$
should be those that most closely approximate these classical basis
states $(p,x)$. According to the uncertainty relation (18), the
existence of a state with exact simultaneous determination of $p$ and
$x$ ($\sigma_{xx} = 0$ and $\sigma_{pp} = 0$) is impossible in quantum
physics. The quantum states that most closely resemble the classical
phase‑space point $(p,x)$ are not eigenstates of position or momentum
alone, but rather the states denoted
$\left| \left.\left\langle z \right\rangle \right\rangle \right.\ $
which saturate the uncertainty relation or more generally the states
denoted
$\left| n,\left. \left\langle z \right\rangle \right\rangle \right. $
\cite{Ranaivoson2022b,Ranaivoson2022,Ravelonjato2023,Ranaivoson2025,Ranaivoson2021}. These latter states are closely related to
the states $\left|\left. \left\langle z \right\rangle \right\rangle \right.\ $
and can be considered as their generalization \cite{Ranaivoson2022}.

\subsection{Joint Momentum-Coordinate states and Quantum Phase Space}

The minimum‑uncertainty states
$\left| \left.\left\langle z \right\rangle \right\rangle \right.\ $,
called joint momentum‑coordinate states form the cornerstone for
defining the QPS, as they provide the closest possible quantum analogue
to a classical point $(p,x)$. The coordinate wavefunction corresponding
to these states are given by the following relation \cite{Ranaivoson2022b,Ranaivoson2022}:

\begin{equation}
\left\langle x \middle| \langle z \rangle \right\rangle = \frac{1}{{(2\pi\mathcal{X)}}^{1/4}}e^{- \frac{\mathcal{B}}{\hslash^{2}}\left( x - \left\langle x \right\rangle \right)^{2} + \frac{i}{\hslash}\left\langle p \right\rangle x}
\end{equation}

in which:

\begin{itemize}
\item $\left\langle p \right\rangle$ and $\left\langle x \right\rangle$ are
  respectively the expectation values of momentum and coordinate
  operators $\mathbf{p}$ and $\mathbf{x}$ corresponding to a state
  $\left| \left.\left\langle z \right\rangle \right\rangle \right.\ $
  itself:

\begin{equation}
\left\langle x \right\rangle = \left\langle \langle z \rangle \middle| \mathbf{x} \middle| \langle z \rangle \right\rangle \quad \text{and} \quad \left\langle p \right\rangle = \left\langle \langle z \rangle \middle| \mathbf{p} \middle| \langle z \rangle \right\rangle
\end{equation}

\item $\mathcal{B}$ is a parameter which is related to the variance and
  covariance, $\sigma_{pp} = \mathcal{P}$, $\sigma_{xx} = \mathcal{X}$ and
  $\sigma_{px} = \mathcal{Q}$ of momentum and coordinate operators,
  $\mathbf{p}$ and $\mathbf{x}$ corresponding to the states
  $\left| \left.\left\langle z \right\rangle \right\rangle \right.\ $:

\begin{equation}
\mathcal{B} = \frac{\hslash^{2}}{4\mathcal{X}}\left(1 - \frac{2i}{\hslash}\mathcal{Q}\right) = \frac{\mathcal{P}}{1 + 4\frac{\mathcal{Q}^{2}}{\hslash^{2}}}\left(1 - \frac{2i}{\hslash}\mathcal{Q}\right)
\end{equation}

and
\begin{equation}
\begin{cases}

\mathcal{P} &= \langle\langle z | (\mathbf{p} - \langle p \rangle)^2 | \langle z \rangle\rangle = \langle\langle z | \mathbf{p}^2 | \langle z \rangle\rangle - \langle p \rangle^2 \\
\mathcal{X} &= \langle\langle z | (\mathbf{x} - \langle x \rangle)^2 | \langle z \rangle\rangle = \langle\langle z | \mathbf{x}^2 | \langle z \rangle\rangle - \langle x \rangle^2 \\
\mathcal{Q} &= \frac{1}{2}\langle\langle z | (\mathbf{p} - \langle p \rangle)(\mathbf{x} - \langle x \rangle) + (\mathbf{x} - \langle x \rangle)(\mathbf{p} - \langle p \rangle) | \langle z \rangle\rangle \\
&= \frac{1}{2}\langle\langle z | \mathbf{px + xp} | \langle z \rangle\rangle - \langle p \rangle \langle x \rangle

\end{cases}
\end{equation}
\end{itemize}

$\mathcal{P}$, $\mathcal{X}$ and $\mathcal{Q}$ correspond to the
saturation of the uncertainty relation (18),

\begin{equation}
\det\left\lbrack \mathcal{G} \right\rbrack = \mathcal{P}\mathcal{X} - \left( \mathcal{Q} \right)^{2} = \frac{(\hslash)^{2}}{4} \quad \text{where} \quad \mathcal{G} = \begin{pmatrix}
\mathcal{P} & \mathcal{Q} \\
\mathcal{Q} & \mathcal{X}
\end{pmatrix}
\end{equation}

$\mathcal{G}$ is the momentum‑coordinate variances‑covariance matrix
corresponding to the state $\left| \left.\left\langle z \right\rangle \right\rangle \right.\ $.
It can be shown that the states $\left| \left.\left\langle z \right\rangle \right\rangle \right.\ $
are eigenstates of the operator $\mathbf{z} = \mathbf{p} - \frac{2i}{\hslash}\mathcal{B} \mathbf{x}$.
The corresponding eigenvalue equation is

\begin{equation}
\mathbf{z}\left| \langle z \rangle \right\rangle = \left\lbrack \left\langle p \right\rangle - \frac{2i}{\hslash}\mathcal{B}\left\langle x \right\rangle \right\rbrack\left| \langle z \rangle \right\rangle = \left\langle z \right\rangle\left| \langle z \rangle \right\rangle
\end{equation}

this eigenvalue equation justifies the notation $\left| \left.\left\langle z \right\rangle \right\rangle \right.\ $

\medskip 

The Quantum Phase Space (QPS) is rigorously defined as the set
{($\left\langle p \right\rangle$, $\left\langle x \right\rangle$)} of
expectation-value pairs ($\left\langle p \right\rangle$,
$\left\langle x \right\rangle$) for given values of the variances and
covariance $\mathcal{P}$, $\mathcal{X}$ and $\mathcal{Q}$ satisfying
(23) \cite{Ranaivoson2022b,Ranaivoson2022}. This is not classical phase space: each QPS point
or centroid ($\left\langle p \right\rangle$,
$\left\langle x \right\rangle$) carries an intrinsic minimum quantum
uncertainty encoded in the covariance matrix $\mathcal{G}$. The states
$\left| \left.\left\langle z \right\rangle \right\rangle \right.\ $
are the most general minimum-uncertainty states. They allow arbitrary
variances $\mathcal{P}$, $\mathcal{X}$ and covariance $\mathcal{Q}$,
corresponding to arbitrary squeezing and rotation of the uncertainty
ellipse in phase space. They are therefore analogous to what are called
squeezed states in the literature \cite{Philbin2014,Bagchi2020,Dodonov2002,RosasOrtiz2019,Dey2018}. The set
$\{\left| \left.\left\langle z \right\rangle \right\rangle \right.\ \}$
is then an overcomplete frame that provides a resolution of the identity
over the QPS \cite{Ranaivoson2022b,Ranaivoson2022}:

\begin{equation}
\iint_{}^{}\left| \langle z \rangle \right\rangle\left\langle \langle z \rangle \right| \frac{d\left\langle p \right\rangle d\left\langle x \right\rangle}{h} = \mathbf{I}
\end{equation}

Any arbitrary instantaneous motional state $\mathbf{\rho}(t)$ of the
particle can then be expressed in the overcomplete frame
$\{\left| \left.\left\langle z \right\rangle \right\rangle \right.\}$

\begin{equation}
\begin{cases}
\displaystyle \mathbf{\rho}(t) = \iint_{}^{}{\iint_{}^{}{\rho(\left\langle z \right\rangle,\left\langle z \right\rangle',t)\left| \langle z \rangle \right\rangle\left\langle \langle z \rangle' \right| \frac{d\left\langle p \right\rangle d\left\langle x \right\rangle}{h}\frac{d\left\langle p \right\rangle'd\left\langle x \right\rangle'}{h}}} \\
\displaystyle \rho(\left\langle z \right\rangle,\left\langle z \right\rangle',t) = \left\langle z(t) \middle| \mathbf{\rho}(t) \middle| z'(t) \right\rangle
\end{cases}
\end{equation}

Here, $h$ is the Planck constant.

\subsection{Time-Dependent orthonormal basis associated to the QPS}

The overcomplete frame $\{\left| \left.\left\langle z \right\rangle \right\rangle \right.\}$
is non-orthogonal: $\left\langle \langle z \rangle \middle| \langle z \rangle' \right\rangle \neq 0$.
Therefore, strict decoherence (the exact vanishing of off‑diagonal
density‑matrix elements) cannot be fully described in this frame. In
fact, as discussed in Section 2, rigorous decoherence requires the
pointer states to be mutually orthogonal so that the reduced density
matrix becomes strictly diagonal. Moreover, the stronger objectivity
criterion of Spectrum Broadcast Structure (SBS) explicitly demands such
orthogonality to guarantee that different environmental fragments
contain perfectly distinguishable information about the system
\cite{Horodecki2015,Korbicz2021}. Since the overcomplete frame
$\{\left| \left. \left\langle z \right\rangle \right\rangle \right. \}$
fails this fundamental orthogonality requirement, the construction of a
second, orthonormal basis should be considered.

To obtain an adequate orthonormal basis, we consider a time-dependent
centroid $\left\langle z \right\rangle = \left\langle z(t) \right\rangle$ and
construct an orthonormal Fock-like basis
$\left\{ \left| n,\left.\left\langle z(\mathbf{t}) \right\rangle \right\rangle \right.\right\}$
above it using ladder operators. The relations between the states
$\left| \left. \left\langle z(t) \right\rangle \right\rangle \right. $
and
$\left| n,\left. \left\langle z(t) \right\rangle \right\rangle \right. $
are analogous to the relations between the ground state of a harmonic
oscillator and its excited states. To formally express these relations,
we introduce ladder operators $\mathbf{Z}$ and
$\mathbf{Z}^{\mathbf{\dagger}}$ which satisfy the following relation
\cite{Ranaivoson2022}:

\vspace{0.25cm}
\begin{equation}
\begin{cases}
\mathbf{Z} = i\frac{\sqrt{\mathcal{X}}}{\hslash}\left( \mathbf{z} - \langle z \rangle \right) = \frac{2\mathcal{B}}{\hslash^{2}}\sqrt{\mathcal{X}}\left[ \left( \mathbf{x} - \langle x \rangle \right) + \frac{i\hslash}{2\mathcal{B}}\left( \mathbf{p} - \langle p \rangle \right) \right] \\
\mathbf{Z}^{\dagger} = - i\frac{\sqrt{\mathcal{X}}}{\hslash}\left( \mathbf{z}^{\dagger} - \langle z \rangle^{*} \right) = \frac{2\mathcal{B}}{\hslash^{2}}\sqrt{\mathcal{X}}\left[ \left( \mathbf{x} - \langle x \rangle \right) - \frac{i\hslash}{2\mathcal{B}}\left( \mathbf{p} - \langle p \rangle \right) \right] \\
\left[ \mathbf{Z},\mathbf{Z}^{\dagger} \right] = \mathbf{Z}\mathbf{Z}^{\dagger} - \mathbf{Z}^{\dagger}\mathbf{Z} = 1
\end{cases}
\end{equation}
\vspace{0.25cm}

We can then consider the following relations, which define the excited
states $\left| n,\left.\left\langle z(t) \right\rangle \right\rangle \right.\ $:

\begin{equation}
\begin{cases}
| n,\langle z \rangle \rangle = \frac{(\mathbf{Z}^{\dagger})^n}{\sqrt{n!}} | \langle z \rangle \rangle \\
\mathbf{Z} | n,\langle z \rangle \rangle = \sqrt{n} | n-1,\langle z \rangle \rangle \\
\mathbf{Z}^{\dagger} | n,\langle z \rangle \rangle = \sqrt{n+1} | n+1,\langle z \rangle \rangle \\
\aleph | n,\langle z \rangle \rangle = \mathbf{Z}^{\dagger}\mathbf{Z} | n,\langle z \rangle \rangle = n | n,\langle z \rangle \rangle
\end{cases}
\end{equation}

In relation (28), $n$ is a positive integer index which labels the
excitation level above the reference state
$\left| \left. \left\langle z \right\rangle \right\rangle \right.\ $.
A state $\left| n,\left. \left\langle z \right\rangle \right\rangle \right.\ $
is an eigenstate of the number operator $\aleph = \mathbf{Z}^{\dagger}\mathbf{Z}$ constructed from
the ladder operators $\mathbf{Z}$ and $\mathbf{Z}^{\dagger}$.
The corresponding eigenvalue is $n$. For a fixed
$\left\langle z \right\rangle = \left\langle z(t) \right\rangle$ (at any
fixed instant $t$), the set $\{\left| n,\left. \left\langle z \right\rangle \right\rangle \right.\}$
form a complete orthonormal basis:

\begin{equation}
\begin{cases}
\displaystyle \sum_{n = 0}^{\infty} | n,\langle z \rangle \rangle \langle n,\langle z \rangle | = \mathbf{I} \\
\langle n,\langle z \rangle | m,\langle z \rangle \rangle = \delta_{nm}
\end{cases}
\end{equation}

The time-dependent orthonormal basis $\{\left| n,\left.\left\langle z(t) \right\rangle \right\rangle \right.\ \}$
encodes the particle's mean trajectory in QPS, with the centroid
$\left\langle z(t) \right\rangle = \left\langle p(t) \right\rangle - \frac{2i}{\hslash}\mathcal{B}(t)\left\langle x(t) \right\rangle$
defining this path. This mean trajectory can be considered as the
quantum analog of a classical trajectory.

Any arbitrary instantaneous motional state $\mathbf{\rho}(t)$ of the
particle can be expressed in the time-dependent orthonormal basis
$\{\left| n,\left. \left\langle z(t) \right\rangle \right\rangle \right.\ \}$
as

\begin{equation}
\begin{cases}
\mathbf{\rho}(t) =\displaystyle \sum_{n = 0}^{\infty}{\sum_{m = 0}^{\infty}{\rho_{m}^{n}(t)}}\left| n,\langle z(t) \rangle \right\rangle \left\langle m,\langle z'(t) \rangle \right| \\
\rho_{m}^{n}(t) = \left\langle n,z(t) \middle| \mathbf{\rho}(t) \middle| m,z(t) \right\rangle
\end{cases}
\end{equation}

\subsection{Identification of Pointer States within the QPS Framework, role of the Hamiltonian and distinction between decoherence and thermalization}

A potential ambiguity arises in the identification of pointer states:
are the pointer states the continuous squeezed states
$\left| \left.\left\langle z \right\rangle \right\rangle \right.\ $
or the time-dependent orthonormal Fock-like states
$\left| n,\left.\left\langle z(t) \right\rangle \right\rangle \right.\ $?
The QPS formalism resolves this by assigning distinct complementary
roles to each, and by showing that all of the main pointer‑state
criteria outlined in Section 2 converge in this unified structure.
We can consider the following two levels:

\begin{itemize}
\item The overcomplete continuous frame $\{\left| \left.\left\langle z \right\rangle \right\rangle \right. \}$
  defines the geometric framework of the QPS. Each state
  $\left| \left.\left\langle z \right\rangle \right\rangle \right.\ $
  corresponds to a localized wave packet centered at
  $(\left\langle p \right\rangle,\left\langle x \right\rangle)$ and
  saturates the uncertainty relation. This framework provides the
  underlying QPS corresponding to the manifold of possible
  classical-like trajectories, i.e., mean trajectories. However, the
  non‑orthogonality of the states
  $\left| \left.\left\langle z \right\rangle \right\rangle \right.\ $
  prevents strict diagonalization of the density matrix.

\item The time-dependent orthonormal basis $\{\left| n,\left. \left\langle z(t) \right\rangle \right\rangle \right.\}$
  can serve as the rigorous pointer basis for strict decoherence,
  enabling diagonalization, although the excited states
  $\left| n,\left.\left\langle z(t) \right\rangle \right\rangle \right.\ $
  do not saturate uncertainty.
\end{itemize}

All criteria from Section 2.2 could converge within this framework:
Einselection, Quantum Darwinism, and Spectrum Broadcast Structure
identify the full orthonormal basis $\{\left| n,\left.\left\langle z(t) \right\rangle \right\rangle \right.\ \}$;
the Predictability Sieve / Minimal Entropy Production single out the
ground state $\left| \left. \left\langle z \right\rangle \right\rangle \right.\ $
as the most classically predictable pure state.

Although the previous discussion shed light on the identification of
pointer states within the QPS framework, the distinction between the two
quantum numbers $\left\langle z \right\rangle$ and $n$ and their
respective roles in decoherence and thermalization can only be fully
understood by examining the structure of the total Hamiltonian.
Consider the general system-bath Hamiltonian (system-environment Hamiltonian)

\begin{equation}
\mathbf{H}_{{SB}} = \mathbf{H}_{{S}} + \mathbf{H}_{{B}} + \mathbf{H}_{{I}}
\end{equation}

where $\mathbf{H}_{\mathbf{S}}$ is the system
self-Hamiltonian, $\mathbf{H}_{\mathbf{B}}$ the bath (environment)
Hamiltonian and $\mathbf{H}_{\mathbf{I}}$ is the interaction
Hamiltonian. A general form of $\mathbf{H}_{\mathbf{I}}$ is given by the
relation (1). Within the QPS framework, we assume that the environment
monitors position, momentum, or both, so the coupling operators
$\mathbf{A}_{k}$ are typically linear combinations of $\mathbf{p}$ and
$\mathbf{x}$:

\begin{equation}
\mathbf{A}_{k} = \alpha_{k}\mathbf{p} + \beta_{k}\mathbf{x}
\end{equation}

As highlighted by relation (2), the pointer states are the approximate
eigenstates of these operators $\mathbf{A}_{k}$, which correspond to
the states $\left| \left. \left\langle z \right\rangle \right\rangle \right. $
and $\left| n,\left. \left\langle z(t) \right\rangle \right\rangle \right. $
as may be expected. It corresponds to the fact that environment thus
continuously measures the centroid
$(\left\langle p \right\rangle,\left\langle x \right\rangle)$.
Decoherence then destroys coherences between different
$\left\langle z \right\rangle$, which correspond to different
classical-like trajectories. The system self-Hamiltonian governs the
evolution related to the excitation number $n$. The key question is
whether $\mathbf{H}_{{S}}$ commutes with the pointer
observables $\mathbf{A}_{k}$:

\begin{itemize}
\item If $\left\lbrack \mathbf{H }_{{S}}, \mathbf {A}_{k} \right\rbrack = 0$, the
  pointer basis $\{\left| n,\left. \left\langle z(t) \right\rangle \right\rangle \right.\}$
  coincides with the energy eigenbasis. In this case, the index $n$ also
  labels energy eigenstates and decoherence destroys coherences between
  different $n$ on the fast timescale $\tau_{D}$. Subsequently,
  thermalization drives the diagonal populations $\rho_{n}^{n}$ toward
  the Gibbs distribution on the slow timescale $\tau_{R}$.

\item If $\left\lbrack \mathbf{H }_{{S}},\mathbf{A}_{k} \right\rbrack \neq 0$, the
  pointer basis $\{\left| n,\left. \left\langle z(t) \right\rangle \right\rangle \right.\}$
  differs from the energy eigenbasis. The environment measures the
  centroid $\left\langle z \right\rangle$ so decoherence destroys
  coherences between different centroids. The system self-Hamiltonian
  couples different $n$ levels and the environment's influence (through
  the interaction) acts on all coherences in the pointer basis,
  including those between different $n$ for the same
  $\left\langle z \right\rangle$, but the rate may depend on $n$.
  Thermalization acts on the diagonal populations $\rho_{n}^{n}$,
  driving them toward the Gibbs distribution.
\end{itemize}

In both cases, the separation of timescales $\tau_{D} \ll \tau_{R}$
arises because decoherence (driven by diffusion) is typically much
faster than thermalization (driven by dissipation). The QPS framework
thus provides a first‑principles geometric distinction between the two
processes.

\subsection{Relations between environment properties, decoherence Regime and the Structure of the QPS}

Based on the results of the previous section, there is a direct
correspondence between the set of pointer states for a particle's motion
and its QPS. Consequently, the characteristics of the environment, which
are reflected in the nature of the pointer states, are directly linked
to the structure of the particle's quantum phase space. This structure
is defined by the variances‑covariance matrix
$\mathcal{G} = \begin{pmatrix}
\mathcal{P} & \mathcal{Q} \\
\mathcal{Q} & \mathcal{X}
\end{pmatrix}$ of the "ground states"
$\left| \left.\left\langle z \right\rangle \right\rangle \right.\ $
defining the QPS. It follows that the properties of the environment
should also be described with these quantities. Two main cases can be
distinguished: stationary and reactive environments.

\begin{itemize}
\item \textbf{Stationary environment (Markovian regime):} this case occurs when
  the environment's intrinsic characteristics, such as its spectral
  density, temperature, and correlation functions, are not significantly
  perturbed or reconfigured by its interaction with the particle.
  Conceptually, the environment is assumed to be so large or have such
  fast internal dynamics that the energy and information exchanged with
  the particle cause negligible back‑action. While the environment
  inevitably captures information about the particle's state (for
  example, through scattering or field emissions), this transfer does
  not substantially alter the environment's global statistical
  properties or its future dynamical rules. Consequently, the
  environment retains no lasting "memory" of any specific interaction
  event; it resets rapidly to its equilibrium state, behaving as a
  memoryless Markovian reservoir. This type of environment corresponds
  precisely to the Markovian limit of decoherence. In this regime, the
  system's dynamics are local in time, described by master equations
  with constant damping rates, such as the Lindblad form. This
  theoretical description is physically justified when the environment's
  correlation time is vanishingly short compared to the system's
  dynamical timescale. The mathematical signature of this limit within
  our quantum phase‑space framework is that the corresponding
  variance‑covariance matrix for the environmental noise is
  time‑independent, reflecting stationary statistical properties.

\begin{equation}
\dot{\mathcal{X}} = \frac{d\mathcal{X}}{dt} = 0, \quad \dot{\mathcal{P}} = \frac{d\mathcal{P}}{dt} = 0, \quad \dot{\mathcal{Q}} = \frac{d\mathcal{Q}}{dt} = 0, \quad \dot{\mathcal{G}} = \frac{d\mathcal{G}}{dt} = \begin{pmatrix}
\dot{\mathcal{P}} & \dot{\mathcal{Q}} \\
\dot{\mathcal{Q}} & \dot{\mathcal{X}}
\end{pmatrix} = 0
\end{equation}

It is this constancy in the second‑order moments that formally
justifies the designation "stationary" for the environment in this
model. This stationarity ensures that the decoherence channels and
diffusion processes they induce are uniform over time, leading to
predictable, exponential decay of coherences.

\item \textbf{Reactive environment (Non‑Markovian regime):} This regime describes
  an environment whose intrinsic physical and statistical properties,
  such as its effective spectral density, temperature distribution, and,
  most crucially, its temporal correlation functions, are notably
  perturbed and reshaped by the particle's presence and dynamics. Unlike
  a passive, Markovian bath that remains inert and unchanging, a
  reactive environment dynamically adapts to the ongoing interaction. It
  does not simply act as a sink for information; rather, the absorption
  of information about the particle's state (e.g., its position,
  momentum, or internal energy) induces a structural or dynamical change
  within the environment itself. This exchange creates a bidirectional
  feedback loop: the system influences the environment, and the
  environment's modified state, in turn, exerts a new influence back on
  the system at a later time. This is the essence of a memory effect;
  the environment effectively retains a non‑trivial, dynamical trace of
  past interactions, which actively influences the future evolution of
  the system's quantum coherences. Consequently, the decoherence process
  becomes history‑dependent. Mathematically, this complex, temporally
  non‑local behavior is captured by a time‑dependent variance‑covariance
  matrix

\begin{equation}
\dot{\mathcal{X}} = \frac{d\mathcal{X}}{dt} \neq 0, \quad \dot{\mathcal{P}} = \frac{d\mathcal{P}}{dt} \neq 0, \quad \dot{\mathcal{Q}} = \frac{d\mathcal{Q}}{dt} \neq 0, \quad \dot{\mathcal{G}} = \frac{d\mathcal{G}}{dt} = \begin{pmatrix}
\dot{\mathcal{P}} & \dot{\mathcal{Q}} \\
\dot{\mathcal{Q}} & \dot{\mathcal{X}}
\end{pmatrix} \neq 0
\end{equation}

This temporal dependence signifies that the statistical moments (the
"noise strength" and inter‑mode correlations) governing the
environmental fluctuations are not stationary but evolve in response to
the system's prior history, providing a direct quantitative signature of
the environment's reactive and memory‑bearing nature. The "breathing" or
rotation of the uncertainty ellipse in phase space provides a direct
geometric representation of non-exponential decoherence, recoherence and
information backflow, corresponding precisely to the non-Markovian
regime.
\end{itemize}

\section{Illustrative application: Extending the Hu-Paz-Zhang model}

This illustrative application demonstrates how the general QPS framework
connects environmental parameters to quantum phase space structure,
naturally accommodating and generalizing the standard HPZ model of
quantum Brownian motion.

\subsection{The Standard Hu‑Paz‑Zhang model}

The Hu--Paz--Zhang (HPZ) master equation is a non-Markovian quantum
master equation that describes the time evolution of a quantum harmonic
oscillator coupled to a general bosonic bath environment (heat bath).
First derived by B. L. Hu, J. P. Paz, and Y. Zhang in 1992 using the
Feynman-Vernon path-integral influence functional method \cite{Feynman1963}, it
serves as a foundational pillar in the study of Quantum Brownian Motion
(QBM), decoherence, and open quantum systems \cite{Breuer2007,deVega2017,Hu1992,Halliwell1996,Homa2023,Homa2020,Rivas2011,Fleming2011,Weiss2021}.

\subsubsection{System‑Bath Hamiltonian}

The model considers a harmonic oscillator of mass $M$ and frequency $\Omega$ linearly coupled to a bath of
harmonic oscillators with masses $m_{k}$ and frequency $\omega_{k}$
which constitutes the environment. The system-bath Hamiltonian is

\begin{equation}
\mathbf{H}_{{SB}} = \mathbf{H}_{{S}} + \mathbf{H}_{{B}} + \mathbf{H}_{{I}} \quad \text{with} \quad 
\begin{cases}
 \mathbf{H}_{{S}} = \dfrac{\mathbf{p}^{2}}{2M} + \dfrac{1}{2}M\Omega^{2}\mathbf{x}^{2} \\[6pt]
\displaystyle \mathbf{H}_{{B}} = \sum_{k} \left( \dfrac{\mathbf{p}_{k}^{2}}{2m_{k}} + \dfrac{1}{2}m_{k}\omega_{k}^{2}\mathbf{q}_{k}^{2} \right) \\[6pt]
\displaystyle \mathbf{H}_{{I}} = \mathbf{x} \sum_{k} C_{k} \mathbf{q}_{k}
\end{cases}
\end{equation}

The bath spectral density corresponding to the coupling with the system
is

\begin{equation}
J(\omega) = \frac{\pi}{2}\sum_{k}^{}{\frac{\left( C_{k} \right)^{2}}{m_{k}\omega_{k}}\delta(\omega - \omega_{k})}
\end{equation}

Since realistic environments contain a very large number of degrees of
freedom, one usually adopts the continuum limit, in which the discrete
set of bath oscillators is replaced by a continuous distribution of
frequencies. Accordingly, the discrete spectral density becomes

\begin{equation}
J(\omega) = \frac{\pi}{2}g(\omega)\frac{\lbrack C(\omega)\rbrack^{2}}{m(\omega)\omega}
\end{equation}

where $g(\omega)$ is the density of bath modes. While $C(\omega)$ and
$m(\omega)$ denote the frequency-dependent oscillator masses and
coupling strengths, respectively.

In the standard HPZ model, only the bath coordinates $\mathbf{q}_{k}$ couple with the system coordinate $\mathbf{x}$, there is no coupling between the bath momenta $\mathbf{p}_{k}$ with the system momentum $\mathbf{p}$. This asymmetry is the defining feature of the standard model and can serve as the starting point for a generalization within the QPS formalism.

\subsubsection{Master equation}

The state of the combined system is described by a density operator $\mathbf{\rho}_{SB}$, evolving via the Liouville‑von
Neumann equation. Tracing over the bath degrees of freedom yields the
reduced density operator (the system density operator) $\mathbf{\rho}_{\mathbf{S}} = \mathbf{\rho}$ which obeys the exact
Hu--Paz--Zhang (HPZ) master equation for quantum Brownian motion in a
general Gaussian environment \cite{Breuer2007,deVega2017}:

\begin{equation}
\frac{d\mathbf{\rho}}{dt} = - \frac{i}{\hslash}\left\lbrack \ \mathbf{H},\mathbf{\rho} \right\rbrack - \frac{i}{\hslash}\Gamma(t)\left\lbrack \mathbf{x,}\left\{ \mathbf{p,\rho} \right\} \right\rbrack - \frac{D_{pp}(t)}{(\hslash)^{2}}\left\lbrack \mathbf{x}\ \mathbf{,}\left\lbrack \mathbf{x,\rho} \right\rbrack \right\rbrack + \frac{D_{xp}(t)}{{(\hslash)}^{2}}\left\lbrack \mathbf{x}\ \mathbf{,\lbrack p,\rho\rbrack} \right\rbrack
\end{equation}

Where

\begin{itemize}
\item $\mathbf{H}$ is the renormalized system Hamiltonian

\begin{equation}
\mathbf{H} = \mathbf{H}_{\mathbf{S}} + \frac{1}{2}M\delta\Omega^{2}(t)\left( \mathbf{x} \right)^{2}
\end{equation}

${\delta\Omega}^{2}(t)$ correspond to a transient time-dependent
frequency renormalization: time‑dependent shift in the effective
oscillation frequency of the Brownian particle, induced by the bath's
back‑reaction. Physically, the particle "drags" a cloud of bath
excitations with it, modifying its effective inertia or potential.

\item $\Gamma(t),\ D_{pp}(t)$ and $D_{xp}(t)$ are respectively the
  friction/dissipation coefficient (driving thermalization), the normal
  diffusion coefficient (driving decoherence), and the anomalous
  diffusion coefficient (coupling position and momentum, modifying
  squeezing). They are determined by the bath spectral density $J(\omega)$ and the bath temperature $T$ \cite{Breuer2007,deVega2017,Hu1992,Homa2023,Rivas2011,Fleming2011}.
\end{itemize}

A widely used example is the Lorentz--Drude spectral density, which
models an Ohmic environment at low frequencies together with a
Lorentzian high-frequency cutoff \cite{Breuer2007,Weiss2021}:

\begin{equation}
J(\omega) = M\gamma_{0}\omega\frac{\omega_{c}^{2}}{\omega_{c}^{2} + \omega^{2}}
\end{equation}

where $\gamma_{0}$ is the damping rate and $\omega_{c}$ the cutoff
frequency. It can be shown in particular that in the Markovian limit
$(\omega_{c} \rightarrow \infty)$, the coefficients reduce to constant
(time-independent) values ($k_{B}$ is the Boltzmann constant):

\begin{equation}
\Gamma(t) \rightarrow \gamma_{0}, \quad D_{pp}(t) \rightarrow 2M\gamma_{0}k_{B}T, \quad D_{xp}(t) \rightarrow 0
\end{equation}

Thus the HPZ master equation reduces to the (Markovian) standard
high-temperature Caldeira-Leggett master equation \cite{Breuer2007,Caldeira1983}:

\begin{equation}
\frac{d\mathbf{\rho}}{dt} = - \frac{i}{\hslash}\left\lbrack \ \mathbf{H},\mathbf{\rho} \right\rbrack - \frac{i}{\hslash}\gamma_{0}\left\lbrack \mathbf{x,}\left\{ \mathbf{p,\rho} \right\} \right\rbrack - \frac{2M\gamma_{0}k_{B}T}{{(\hslash)}^{2}}\left\lbrack \mathbf{x}\ \mathbf{,\lbrack x,\rho\rbrack} \right\rbrack
\end{equation}

Beyond the Lorentz--Drude model, different choices of spectral density
allow one to describe different environmental properties, including
Ohmic, sub-Ohmic, super-Ohmic, and structured (colored) reservoirs,
leading to different memory and dissipation effects \cite{Breuer2007,deVega2017,Weiss2021}.

\subsection{Extension of the HPZ model within the QPS formalism}

The standard HPZ master equation is fully consistent with the QPS
formalism. It can provide explicit relations between the
variance-covariance matrix $\mathcal{G}(t)$ of the QPS ground states and
the HPZ coefficients, and it can be formulated in both the overcomplete
frame $\{\left| \left. \left\langle z \right\rangle \right\rangle \right.\}$
and the orthonormal pointer basis $\{\left| n,\left.\left\langle z(t) \right\rangle \right\rangle \right. \}$.
However, while the standard HPZ model is sufficient for environments
that monitor only the system's position, the QPS formalism treats
position and momentum as complementary phase-space variables. It is
therefore \textbf{more natural}, and particularly relevant for nanoscience,
to consider the extended version of the HPZ model that includes momentum
coupling.

\subsubsection{Extended System‑Bath Hamiltonian}

The standard HPZ Hamiltonian couples the bath only to the particle's position $\mathbf{x}$ as shown
in the relation (35). To describe environments that monitor both
position and momentum, a situation common in nanoscale devices where
multiple decoherence channels are active, the interaction Hamiltonian
must be modified to include momentum coupling. The inclusion of
momentum-dependent couplings in quantum Brownian motion has been
explored in recent studies, where the standard HPZ equation emerges as a
special case \cite{Huang2022,Ferialdi2017,Bhattacharjee2024}. The present work differs from these prior
generalizations in both motivation and scope. While previous studies
focused on deriving generalized master equations, our approach is rooted
in the Quantum Phase Space (QPS) formalism and aims to illustrate its
power by explicitly linking environmental parameters, encoded in the
spectral-density matrix, to the QPS geometry and the emergence of
classical-like trajectories. Furthermore, the QPS formalism naturally
accommodates both Markovian and non-Markovian regimes through the
time-dependence of the covariance matrix $\mathcal{G}(t)$, providing a
unified geometric language that is absent in previous momentum-coupling
generalizations. The most general interaction Hamiltonian including this
coupling and which is naturally compatible with the QPS framework (cf.
Eq. (2)) is

\begin{equation}
\mathbf{H}_{\mathbf{I}} = \mathbf{p}\sum_{\mathbf{k}}^{}{\left( A_{k}\mathbf{q}_{\mathbf{k}} + B_{k}\mathbf{p}_{\mathbf{k}} \right)} + \mathbf{x}\sum_{\mathbf{k}}^{}{\left( C_{k}\mathbf{q}_{\mathbf{k}} + D_{k}\mathbf{p}_{\mathbf{k}} \right)}
\end{equation}

where the system momentum $\mathbf{p}$ and position $\mathbf{x}$ are
allowed to couple linearly to both the bath coordinates
$\mathbf{q}_{k}$ and bath momenta $\mathbf{p}_{k}$.
This general interaction encompasses the standard HPZ model as a
particular case $(A_{k} = 0,\ B_{k} = 0,\ D_{k} = 0)$ while also
describing momentum-mediated and mixed coupling mechanisms that may
arise in quantum transport, kinetic interactions, or engineered quantum
reservoirs.

Unlike the standard HPZ model, which is completely characterized by a
single spectral density [Eq. (36)], the generalized interaction
requires a spectral-density matrix

\begin{equation}
J(\omega) = \begin{pmatrix}
J_{pp}(\omega) & J_{px}(\omega) \\
J_{xp}(\omega) & J_{xx}(\omega)
\end{pmatrix}
\end{equation}

whose elements describe the correlations associated with the different
coupling channels. For the discrete bath, the spectral densities are
defined by

\begin{equation}
\begin{cases}
J_{pp}(\omega) = \dfrac{\pi}{2}\sum_{k} \left[ \dfrac{A_{k}^{2}}{m_{k}\omega_{k}} + m_{k}\omega_{k} B_{k}^{2} \right] \delta(\omega - \omega_{k}) \\[8pt]
J_{xx}(\omega) = \dfrac{\pi}{2}\sum_{k} \left[ \dfrac{C_{k}^{2}}{m_{k}\omega_{k}} + m_{k}\omega_{k} D_{k}^{2} \right] \delta(\omega - \omega_{k}) \\[8pt]
J_{px}(\omega) = \dfrac{\pi}{2}\sum_{k} \left[ \dfrac{A_{k}C_{k}}{m_{k}\omega_{k}} + m_{k}\omega_{k} B_{k}D_{k} \right] \delta(\omega - \omega_{k}) \\[8pt]
J_{xp}(\omega) = J_{px}(\omega)
\end{cases}
\end{equation}

As in the standard HPZ model, the bath is assumed to contain a
macroscopic number of oscillators. In the continuum (thermodynamic)
limit, the discrete frequency spectrum is replaced by a continuous
distribution of modes, and the spectral densities become smooth
functions of the frequency,

\begin{equation}
\begin{cases}
J_{pp}(\omega) = \dfrac{\pi}{2}g(\omega)\left[ \dfrac{A(\omega)^{2}}{m(\omega)\omega} + m(\omega)\omega B(\omega)^{2} \right] \\[8pt]
J_{xx}(\omega) = \dfrac{\pi}{2}g(\omega)\left[ \dfrac{C(\omega)^{2}}{m(\omega)\omega} + m(\omega)\omega D(\omega)^{2} \right] \\[8pt]
J_{px}(\omega) = \dfrac{\pi}{2}g(\omega)\left[ \dfrac{A(\omega)C(\omega)}{m(\omega)\omega} + m(\omega)\omega B(\omega)D(\omega) \right] \\[8pt]
J_{xp}(\omega) = J_{px}(\omega)
\end{cases}
\end{equation}

where $g(\omega)$ is the density of bath modes.

\subsubsection{Generalized Master equation}

By performing a partial trace over the environmental degrees of freedom, we obtain the system's density operator $\mathbf{\rho}$ which satisfies the following generalized non-Markovian master equation

\begin{equation}
\frac{d\mathbf{\rho}}{dt} = - \frac{i}{\hslash}\left\lbrack \ \mathbf{H},\mathbf{\rho} \right\rbrack - \frac{i}{\hslash}\sum_{i,j = p,\ x}^{}{\Gamma_{ij}(t)}\left\lbrack \mathbf{q}_{i}\mathbf{,}\left\{ \mathbf{q}_{j}\mathbf{,\rho} \right\} \right\rbrack - \frac{1}{{(\hslash)}^{2}}\sum_{i,j = p,\ x}^{}{D_{ij}(t)}\left\lbrack \mathbf{q}_{i}\mathbf{,\lbrack}\mathbf{q}_{j}\mathbf{,\rho\rbrack} \right\rbrack
\end{equation}

Where

\begin{itemize}
\item $\mathbf{q}_{p} = \mathbf{p}$ and $\mathbf{q}_{x} = \mathbf{x}$
\item $\mathbf{H}$ is the renormalized system Hamiltonian

\begin{equation}
\mathbf{H} = \mathbf{H}_{\mathbf{S}} + \frac{1}{2}\delta\Omega^{2}(t)\mathbf{x}^{2} - \frac{\delta M(t)}{2M}\mathbf{p}^{2} + \frac{\delta C(t)}{2}(\mathbf{xp + px})
\end{equation}

${\delta\Omega}^{2}(t),\ \delta M(t)$ and $\delta C(t)$ denote the
time-dependent renormalizations of the oscillator frequency, effective
mass, and position-momentum coupling induced by the interaction with the
environment.

\item The coefficients $\Gamma_{ij}(t)$ constitute the dissipation
  matrix, whereas $D_{ij}(t)$ form the diffusion matrix. Their
  diagonal elements describe the dissipation and diffusion associated
  with the position and momentum coupling channels, whereas the
  off-diagonal elements account for the correlations induced by the
  mixed system-bath interactions. All these coefficients are uniquely
  determined by the bath temperature $T$ and the spectral-density matrix
  $J_{ij}(\omega)$ through the corresponding dissipation and noise
  kernels.
\end{itemize}

\subsubsection{Expectation values and Variance-Covariance}

As we have for the expectation values and variance-covariance of position and momentum the following relations

\begin{equation}
\begin{cases}
\left\langle p \right\rangle = \operatorname{Tr}\left( \mathbf{\rho p} \right) \\
\left\langle x \right\rangle = \operatorname{Tr}\left( \mathbf{\rho x} \right) \\
\mathcal{P} = \operatorname{Tr}\left\lbrack \mathbf{\rho}\left( \mathbf{p -}\left\langle p \right\rangle \right)^{2} \right\rbrack \\
\mathcal{X} = \operatorname{Tr}\left\lbrack \mathbf{\rho}\left( \mathbf{x -}\left\langle x \right\rangle \right)^{2} \right\rbrack \\
\mathcal{Q} = \frac{1}{2}\operatorname{Tr}\left\lbrack \mathbf{\rho}\left( \mathbf{px + xp} \right) \right\rbrack - \left\langle p \right\rangle\left\langle x \right\rangle
\end{cases}
\end{equation}

It can be deduced from the general master equation (47), that we have
the following relation for the evolution of the centroid
$(\left\langle p(t) \right\rangle,\left\langle x(t) \right\rangle)$
corresponding to the states $\left| n,\left. \left\langle z(t) \right\rangle \right\rangle \right.\ $
and for the ground variance-covariance matrix $\mathcal{G} = \begin{pmatrix}
\mathcal{P} & \mathcal{Q} \\
\mathcal{Q} & \mathcal{X}
\end{pmatrix}$:

\begin{equation}
\begin{cases}
\dfrac{d\langle p \rangle}{dt} = - [\delta C(t) + 2\Gamma_{xp}(t)]\langle p \rangle - [M( \Omega^{2} + \delta\Omega^{2}(t)) + 2\Gamma_{xx}(t)]\langle x \rangle \\[4pt]
\dfrac{d\langle x \rangle}{dt} = \left[\dfrac{1 - \delta M(t)}{M} - 2\Gamma_{pp}(t)\right]\langle p \rangle + [\delta C(t) - 2\Gamma_{px}(t)]\langle x \rangle \\[4pt]
\dfrac{d\mathcal{P}}{dt} = - 2[\delta C(t) + 2\Gamma_{xp}(t)]\mathcal{P} - 2[M( \Omega^{2} + \delta\Omega^{2}(t)) + 2\Gamma_{xx}(t)]\mathcal{Q} + 2D_{pp} \\[4pt]
\dfrac{d\mathcal{X}}{dt} = 2[\delta C(t) - 2\Gamma_{px}(t)]\mathcal{X} + 2\left[\dfrac{1 - \delta M(t)}{M} - 2\Gamma_{pp}(t)\right]\mathcal{Q} + 2D_{xx} \\[4pt]
\dfrac{d\mathcal{Q}}{dt} = \left[\dfrac{1 - \delta M(t)}{M} - 2\Gamma_{pp}(t)\right]\mathcal{P} - [M( \Omega^{2} + \delta\Omega^{2}(t)) + 2\Gamma_{xx}(t)]\mathcal{X} - 2[\Gamma_{px}(t) + \Gamma_{xp}(t)]\mathcal{Q} + 2D_{xp}
\end{cases}
\end{equation}

The equations in (50) correspond to the general non-Markovian regime
(reactive environment) as described by Eq. (34). These equations
establish a direct link between the microscopic parameters of the
extended HPZ model and the QPS geometry: the centroid dynamics describe
classical-like trajectories influenced by the environment's memory,
while the variance-covariance equations govern the evolution and
time-dependent deformation of the ground uncertainty ellipse. In the Markovian
limit (stationary environment, Eq. 33), the coefficients of the master
equation become constant and we have the following reduction

\begin{equation}
\Gamma_{ij}(t) \longrightarrow \Gamma_{ij}^{(0)}, \quad D_{ij}(t) \longrightarrow D_{ij}^{(0)}, \quad \delta\Omega^{2}(t) = 0, \quad \delta M(t) = 0, \quad \delta C(t) = 0
\end{equation}

The equations in (50) become

\begin{equation}
\begin{cases}
\dfrac{d\langle p \rangle}{dt} = - 2\Gamma_{xp}^{(0)}\langle p \rangle - \left( M\Omega^{2} + 2\Gamma_{xx}^{(0)} \right)\langle x \rangle \\[4pt]
\dfrac{d\langle x \rangle}{dt} = \left(\dfrac{1}{M} - 2\Gamma_{pp}^{(0)}\right)\langle p \rangle - 2\Gamma_{px}^{(0)}\langle x \rangle \\[4pt]
2\Gamma_{xp}^{(0)}\mathcal{P} + (M\Omega^{2} + 2\Gamma_{xx}^{(0)})\mathcal{Q} = D_{pp} \\[4pt]
2\Gamma_{px}^{(0)}\mathcal{X} - \left( \dfrac{1}{M} - 2\Gamma_{pp}^{(0)} \right)\mathcal{Q} = D_{xx} \\[4pt]
- \left( \dfrac{1}{M} - 2\Gamma_{pp}^{(0)} \right)\mathcal{P} + \left( M\Omega^{2} + 2\Gamma_{xx}^{(0)} \right)\mathcal{X} + 2(\Gamma_{px}^{(0)} + \Gamma_{xp}^{(0)})\mathcal{Q} = - 2D_{xp}
\end{cases}
\end{equation}

The equations in (52) correspond to the Markovian regime (stationary
environment), where all coefficients have become time-independent and
the transient renormalizations vanish. In this limit, the centroid
dynamics reduce to a system of linear differential equations with
constant coefficients, describing damped motion of the mean trajectory.
The ground variance-covariance equations, however, become algebraic
relations rather than differential equations, reflecting the
stationarity of the environment. These algebraic relations define the
fixed ground uncertainty ellipse in phase space: they explicitly link the
constant diffusion and dissipation coefficients to the time-independent
QPS geometry, formalizing the concept that a stationary environment
selects and "freezes" a specific set of quasi-classical trajectories.
For the Markovian high temperature limit of the standard HPZ model:
$\Gamma_{xx}^{(0)} = \Gamma_{pp}^{(0)} = \Gamma_{px}^{(0)} = 0,\ \Gamma_{xp}^{(0)} = \gamma_{0}$
and $D_{px} = 0,\ D_{xx} = 0,\ D_{pp} = 2M\gamma_{0}k_{B}T$, we obtain

\begin{equation}
\begin{cases}
\dfrac{d\left\langle p \right\rangle}{dt} = - 2\gamma_{0}\left\langle p \right\rangle - M\Omega^{2}\left\langle x \right\rangle \\[4pt]
\dfrac{d\left\langle x \right\rangle}{dt} = \dfrac{\left\langle p \right\rangle}{M} \\[4pt]
\mathcal{P} = Mk_{B}T \\[4pt]
\mathcal{Q} = 0 \\[4pt]
\mathcal{X} = \dfrac{\mathcal{P}}{M^{2}\Omega^{2}} = \dfrac{k_{B}T}{M\Omega^{2}}
\end{cases}
\end{equation}

According to equations in (53), the centroid dynamics reduce to
classical-like damped oscillator equations. The equilibrium QPS geometry
satisfies the equipartition theorem, with zero covariance
$(\mathcal{Q} = 0)$ indicating an aligned, non-squeezed ground uncertainty ellipse. These results confirm that the QPS framework consistently
reproduces the expected thermal equilibrium state of a harmonic
oscillator.

\subsection{Pointer States}

The generalized interaction Hamiltonian (42) simultaneously couples both the position and momentum observables of the system to the environment. Since the QPS formalism is likewise constructed on the simultaneous treatment of coordinates and momenta, it is natural to expect that the pointer states are the states $\left| \left.\left\langle z \right\rangle \right\rangle \right.\ $ and $\left| n,\left.\left\langle z(t) \right\rangle \right\rangle \right.\ $ introduced within the QPS framework as discussed in Section 4.4. However, a rigorous demonstration of this conjecture would require:

\begin{itemize}
\item Solving the generalized master equation (47) in the proposed QPS pointer-state basis,
\item Analyzing the corresponding decoherence rates for the off-diagonal elements,
\item Verifying explicitly that the pointer basis is indeed the proposed ones.
\end{itemize}

Such an analysis is beyond the scope of the present work because of the lengthy calculations involved, but it constitutes a promising direction for future research. The generalized HPZ master equation derived here provides the necessary dynamical framework for such an investigation. Beyond its theoretical interest, this generalized master equation may provide a suitable framework for modeling decoherence in nanoscale systems where both position and momentum interactions with the environment play a significant role. Potential applications include:

\begin{itemize}
\item Nanoelectromechanical resonators: Position and momentum dependent coupling to quantum point contacts and other mesoscopic detectors \cite{Seoanez2007,Ablimit2021,Cleland2024,Bozkurt2025,Shevyrin2024,Vaidya2017}
\item Optomechanical systems: Radiation-pressure interactions, photon recoil, and structured optical environments \cite{Schafer2024,Noguchi2020,Zhou2022}
\item Superconducting quantum circuits: Flux, charge, and dielectric noise \cite{Heinrich2021,Bhattacharjee2024,Burkard2004,Marchegiani2025}
\item Semiconductor quantum dots: Charge noise and electron-phonon coupling \cite{Hayashi2003,Ablimit2021,Ranni2024,Ye2024}
\item Hybrid quantum devices: Multiple decoherence channels and engineered environments \cite{Lacroix2024,Brandner2025,Lastra2011,Wu2025,Youssefi2023,Chia2024}
\end{itemize}

In these systems, non-Markovian effects and phase-space dynamics are known to influence coherence times and quantum control. Within these systems, the QPS formalism may offer an alternative representation of decoherence dynamics based directly on the evolution of the phase-space moments, as given by equations (50). The generalized master equation with its time-dependent coefficients naturally captures the non-Markovian features of these nanoscale environments, while the QPS geometry provides a clear physical interpretation of the decoherence dynamics and regimes in terms of the evolution of the uncertainty ellipse.

\section{Discussion and Conclusion}

This work has introduced and developed a novel theoretical framework grounded in the concept of Quantum Phase Space (QPS) to address the fundamental and practical challenges of quantum decoherence in nanoscience. Our approach unifies the dual pillars of the decoherence process: the dynamical selection of pointer states and the characterization of decoherence regimes. By connecting environmental properties to the geometry of quantum phase space, we provide a powerful tool for modeling, analyzing, and potentially controlling quantum coherence at the nanoscale.

Our central achievement is the identification of potential pointer states for a particle's motional degrees of freedom as minimum-uncertainty states $\left| \left. \left\langle z \right\rangle \right\rangle \right.\ $ that saturate the uncertainty relation along with their "excited" counterparts $\left| n,\left.\left\langle z(t) \right\rangle \right\rangle \right.\ $. These states are directly related to the concept of QPS and resolve the theoretical impasse presented by idealised position or momentum eigenstates, which are physically inadmissible under the uncertainty principle. The states $\left| \left. \left\langle z \right\rangle \right\rangle \right.\ = \left| 0,\left.\left\langle z \right\rangle \right\rangle \right.\ $ (the "ground states" defining the QPS) and their excitations $\left| n,\left.\left\langle z(t) \right\rangle \right\rangle \right. $ constitute the closest quantum analogue to a classical phase‑space point. Consequently, they provide the natural, physically sound potential pointer basis into which a system (particle in motion) decoheres \cite{Schlosshauer2004,Zurek2003,Schlosshauer2007,Ranaivoson2022b,Wheeler1983,Manampisoa2025,Zurek1991,Brasil2015,Duruisseau2023,Chisholm2026,Zurek2022,Feller2020,Zurek1993a,Zurek1993b,Dalvit2005}.

The structure of the QPS is encoded in the variance‑covariance matrix $\mathcal{G} = \begin{pmatrix} \mathcal{P} & \mathcal{Q} \\ \mathcal{Q} & \mathcal{X} \end{pmatrix}$ of these ground states $\left| \left.\left\langle z \right\rangle \right\rangle \right.\ $. We have seen that this matrix is not static but is directly shaped by the environment, providing a unifying geometric criterion to distinguish decoherence regimes:

\begin{itemize}
\item \textbf{Markovian (stationary) regime:} This corresponds to a memoryless, broadband environment whose statistical properties are unchanging. Within our framework, this is characterized by a time-independent variance-covariance matrix $(\dot{\mathcal{G}} = 0)$. In the Markovian limit of the extended HPZ model that we introduce in this paper, we derive the algebraic relations (51), which explicitly link the constant diffusion and dissipation coefficients to the fixed uncertainty ellipse in phase space. This formalizes the concept that a stationary environment selects and "freezes" a specific set of quasi-classical trajectories \cite{Manampisoa2025,Isar1999}.

\item \textbf{Non‑Markovian (reactive) regime}: This describes environments with memory, structure, or strong coupling, where past interactions influence future dynamics. Here, the environment is reactive, and its statistical imprint on the system evolves. Mathematically, this is captured by a time-dependent variance-covariance matrix $(\dot{\mathcal{G}} \neq 0)$. Using the generalized non-Markovian master equation derived from our extended HPZ model, it is shown how time-dependent diffusion and friction coefficients govern the evolution of $\mathcal{G}(t)$ via the generalized equations (49). The resulting "breathing" or rotation of the uncertainty ellipse in phase space provides a direct geometric representation of non‑exponential decoherence, recoherence, and information backflow \cite{Breuer2009,deVega2017,Breuer2016,Malekakhlagh2016,Lacroix2024,Brandner2025,Odeh2025,Rivas2010,Breuer2012,Li2012,Piilo2008,Fanchini2014,Wu2025,Ferialdi2017}.
\end{itemize}

\medskip

The extended interaction Hamiltonian (43), which gives rise to the spectral-density matrix (44) and the generalized HPZ master equation (47), accounts for all possible bilinear couplings between the system and the bath. From this generalized master equation, we derive explicit closed-form evolution equations for the centroid $(\left\langle p(t) \right\rangle,\left\langle x(t) \right\rangle)$ of the classical-like trajectories and the ground variance-covariance matrix $\mathcal{G}(t)$ directly linking the environmental parameters to the QPS geometry. This approach provides a suitable framework for modeling decoherence in nanoscale systems where both position and momentum interactions with the environment play a significant role, including nanoelectromechanical resonators \cite{Seoanez2007,Ablimit2021,Cleland2024,Bozkurt2025,Shevyrin2024,Vaidya2017}, optomechanical systems \cite{Schafer2024,Noguchi2020,Zhou2022}, superconducting quantum circuits \cite{Heinrich2021,Bhattacharjee2024,Burkard2004,Marchegiani2025}, semiconductor quantum dots \cite{Hayashi2003,Ablimit2021,Ranni2024,Ye2024} and hybrid quantum devices \cite{Lacroix2024,Brandner2025,Lastra2011,Wu2025,Youssefi2023,Chia2024}.

\medskip 

A key strength of the QPS framework is its ability to seamlessly encompass both Markovian and non‑Markovian dynamics within a single, consistent mathematical language. It transcends the traditional modelling dichotomy by showing that both regimes are manifestations of how environmental properties, encoded in spectral densities and correlation functions, imprint themselves onto the quantum phase space geometry. The QPS framework offers a clear and simple metric (the time‑dependence of $\mathcal{G}(t)$) to quantify non‑Markovian memory. This can guide the active harnessing of environmental back action for quantum advantage, such as using structured environments for enhanced quantum sensing or leveraging memory effects in reservoir engineering for information processing \cite{Breuer2007,deVega2017,Breuer2016,Rivas2011,Fleming2011}.

\medskip

Beyond its implications for nanoscience and quantum technologies, the QPS framework developed here has profound significance for fundamental physics and the foundational problems of quantum mechanics. The identification of pointer states as the natural basis for decoherence addresses the quantum measurement problem by providing a rigorous, geometric mechanism for the emergence of classical behavior from quantum dynamics \cite{Schlosshauer2004,Zurek2003,Schlosshauer2007,Ranaivoson2022b,Wheeler1983,Manampisoa2025,Tomaz2025,Karlsson2025}. Moreover, the QPS formalism is not restricted to non-relativistic quantum mechanics. Recent developments have extended the concept of quantum phase space to the relativistic domain, establishing a relativistic quantum phase space that inherently incorporates both the uncertainty principle and relativistic covariance \cite{Ranaivoson2022,Ravelonjato2023,Ranaivoson2025,Ranaivoson2021,Randriantsoa2026,Rasamimanana2026,Randriantsoa2025}. These developments suggest that the QPS framework may also provide a relativistic geometric framework for studying decoherence processes in contexts where relativistic effects become important. The relevance of such a relativistic study of decoherence spans nanoscale physics, fundamental physics, and cosmology. At the nanoscale, mechanical oscillators in quantum states provide sensitive probes of quantum gravitational effects, where fluctuating spacetime induces momentum-basis decoherence \cite{Donadi2025}, and gravitational time dilation itself can cause universal decoherence of composite particle superpositions \cite{Pikovski2015}. Furthermore, open quantum dots in graphene offer a unique platform for studying relativistic pointer states, as graphene's Dirac fermions imitate massless relativistic particles \cite{Ferry2010}. At the most fundamental level, the Unruh effect plays a dual role as both a decoherence source and an enhancer of quantum correlations for accelerating quantum systems \cite{Shang2025}, while black holes and Killing horizons have been shown to decohere quantum superpositions through the emission of extremely low-frequency Hawking quanta \cite{Danielson2025}. In cosmology, decoherence is essential for understanding the quantum-to-classical transition of primordial curvature perturbations during inflation \cite{Sano2025}, with the superhorizon environment providing the mechanism for classical density perturbations to emerge from quantum fluctuations. These diverse contexts, from laboratory-scale tabletop experiments testing gravity-induced decoherence \cite{Bose2025} to the cosmological emergence of classical structure, all require the kind of unified, geometric decoherence framework that QPS provides.

\medskip 

The connection between the generalized HPZ model and the QPS formalism suggests that the pointer states identified within the QPS framework naturally emerge from the decoherence dynamics. A rigorous demonstration of this conjecture, solving the generalized master equation in the QPS pointer-state basis and analyzing the corresponding decoherence rates, is left for future work. However, the framework established here provides the necessary foundation for such an investigation.

\medskip

In conclusion, our Quantum Phase Space framework turns the abstract problem of decoherence into a concrete, visual, and geometrically sound approach. By integrating the dynamics of pointer state selection and coherence decay into the evolving structure of the quantum phase space, this approach provides a powerful synthetic tool. It not only deepens our fundamental understanding of how quantum systems interact with their environments, addressing foundational questions about the quantum-to-classical transition and the measurement problem, but also may deliver a practical and unified methodology for modeling, mitigating, and potentially harnessing decoherence, particularly in the fields of nanoscience and quantum technologies. Furthermore, the relativistic extension of the QPS framework, with its symmetry group based on Linear Canonical Transformations and its connections to particle physics and cosmology suggests that the geometric approach developed here may have far-reaching implications across physics, spanning the subatomic, nanoscopic, and cosmological scales.

\section*{Declarations}

\subsection*{Conflict of Interest}

The authors declare that they have no known competing financial interests or personal relationships that could have appeared to influence the work reported in this paper.

\subsection*{Use of Artificial Intelligence (AI)}

During the preparation of this manuscript, the authors used AI to improve the English language and grammar, as well as to assist with reference formatting, organization, and verification. No AI tool was used to generate scientific data, results, or conclusions. The authors are fully responsible for all published content.

\subsection*{Acknowledgements}

The authors would like to thank the Institut National des Sciences et Techniques Nucléaires (INSTN) for organizing the NANOMADA 2025 colloquium, where a part of this work was presented.

\subsection*{Funding}

This research did not receive any specific grant from funding agencies in the public, commercial, or not-for-profit sectors.

\newpage


\end{document}